\documentclass[a4paper,12pt]{article}
\pdfoutput=1
\usepackage{jheppub}
\usepackage[utf8]{inputenc}
\usepackage{amsmath}
\title{\boldmath Wilson Loops for  M2- and M5-brane spaces }

\author{Edward Quijada and Henrique Boschi-Filho}

\affiliation{Instituto de F\'\i sica, Universidade Federal de Rio de Janeiro,\\Caixa Postal 68528, RJ 21941-972, Brazil}

\emailAdd{edward@if.ufrj.br}
\emailAdd{boschi@if.ufrj.br}

\abstract{We calculate configuration energies of string-like membranes in M2- and M5-brane spaces. In the near horizon approximation these backgrounds reduce to $AdS_4\times S^7$ and $AdS_7\times S^4$ spaces and the dual theories are supersymmetric $SU(N)$ gauge theories, in accordance with the AdS/CFT correspondence. In this case the string-like configuration energy is identified with the quark and anti-quark energy interaction in the dual theories.  
Far from the horizon the dual theory is more involved but any how we were able to calculate the string-like configuration energy. 
For limiting regimes we find simpler solutions for which 
some potentials exhibit a confinement behavior while others are non-confining.}

\begin{document}
\maketitle
\flushbottom
\section{Introduction}

The AdS/CFT correspondence is a duality between string/M theory in $AdS_{n+1}\times S^p$ and
supersymmetric conformal $SU(N)$ Yang-Mills theories, for large $N$, in $n$-dimensional flat space. 
This correspondence was proposed by Maldacena \cite{maldacena1} and soon after detailed
by Gubser, Klebanov and Polyakov \cite{GKP} and Witten \cite{Witten1}. For a review,
see for instance \cite{Review1,Review2}. 

After the proposal of the AdS/CFT correspondence, 
Maldacena \cite{maldacena2} and independently Rey and Yee \cite{Rey:1998ik}  (MRY)
proposed a method to calculate expectation values of an 
operator similar to the Wilson loop for the large $N$ limit of field theories.  
The Wilson loop operator is defined as $W(C)=\dfrac{1}{N}TrPe^{i\oint_{C}A}$, 
where $C$ denotes a closed loop in space-time and the trace is over
the fundamental representation of the gauge field $A$. In the particular case of a rectangular 
loop (of sides $T$ and $L$), it is possible to calculate 
(in the limit $T\rightarrow\infty$) the expectation value for the Wilson loop: 
$<W(C)>=A(L)e^{-TE(L)}$, where $E(L)$ is the energy of the quark-antiquark pair. 
The MRY proposal claims that the expectation value of the Wilson loop correspond 
to the worldsheet area $S$ of a string whose boundary is the loop in 
question i.e. $<W(C)>\sim e^{S}$.
In this way, Maldacena calculated the quark-antiquark potential for the case of 
$AdS_5\times S^5$. 
The result is a non-confining potential for the quark-antiquark interaction, which is 
consistent with the conformal symmetry of the dual super Yang-Mills theory. 
A discussion of the Wilson loops for strings on a certain class of curved 10 dimensional spacetimes 
was presented in \cite{kinar}, where a confinement criterion was obtained. 

A finite temperature calculation of the Wilson loops following the MRY approach was
given in \cite{finiteT,brandhuber} by considering an AdS Schwarzschild background, 
where the temperature of the conformal dual theory is identified with the Hawking 
temperature of the black hole, as in the original Witten's work \cite{Witten2}. 
This set up also behaves as a non-confining potential for the quark-antiquark 
interaction.

An interesting modification of the zero temperature set up is to consider the string in a $D3$-brane background. 
This case was treated in \cite{henrique}, where different behaviors for the potentials were obtained, 
some of them confining and others not. This is understood based on the fact that the $D3$-brane does not have 
conformal symmetry, but in the near horizon approximation it reproduces asymptotically the $AdS$ space. 

A simpler situation was considered calculating the Wilson loop for the string in some phenomenological  
holographic AdS/QCD models. In the case of the hard-wall model, a confining behavior was obtained 
\cite{henrique2,Andreev:2006ct} with a match with the Cornell potential.
A finite temperature version of this calculation was given in \cite{henrique3}, 
where a second order phase transition takes place describing qualitatively a 
confinement/deconfinement phase transition. 
Actually, it was shown that a Hawking-Page phase transition \cite{HP} occurs for the hard- and 
soft-wall models at finite temperature 
\cite{Herzog,henrique4,Andreev:2006eh,Andreev:2006nw,Kajantie:2006hv}
and in the case of the soft-wall model, a good estimate of the deconfinement temperature, 
compatible with QCD predictions, was found \cite{Herzog}. 

In this work, we start from bosonic membranes moving in spaces generated by  $N$
$M2$-branes and $M5$-branes. In order to investigate the behavior 
of the Wilson loops in this set up, we need first to derive a string action from a membrane set-up.
This is made by compactification of one world-volume coordinate simultaneously with 11-dimension of the background spaces (see \cite{duff}). Then the membrane assumes a cigar or string-like shape if the compatified dimension is small enough compared to the size of the other dimensions.  Following the MRY approach \cite{maldacena2,Rey:1998ik}, we compute the energy of these configurations. In the cases where a dual theory can be found, this energy is identified with the potencial interaction between heavy quarks attached to the ends of the string. 
The solution for the quark-antiquark potential in these spaces are expressed as integrals involving metric elements. We consider different limits for these potentials 
with highly curved or almost straight geodesics as well as far from or close to the branes
approximations. In some limiting cases we obtain a confining behavior. 
This work can be understood as an extension of the calculation of Wilson loops \cite{maldacena2,Rey:1998ik} to the case of $M2$- and $M5$-brane backgrounds. In \cite{maldacena2} Maldacena discussed the case of M5-branes using the Wilson surface, which is a different object. A connection between M2-branes and Wilson loops have also appeared recently in \cite{Farquet:2014bda} with a different perspective. 

Another motivation for this study is the possibility of application in two-dimensional condensed matter systems like the ones showing superconductivity or quantum Hall effect for the case of M2-branes since the dual theory when it exists is defined on a (2+1) dimensional space. Usually, these applications appear connected with the $AdS_{4}\times S^7$ space (for reviews, see for instance  \cite{Hartnoll:2009sz,Herzog:2009xv,McGreevy:2009xe,Horowitz:2010gk}) modified to include a black hole in order to describe finite temperature systems. The M5-brane space may be relevant to discuss six-dimensional gauge theories \cite{Seiberg:1996vs,Witten:1997kz}.


\subsection{Membrane formulae}

The bosonic part for M-theory supermembrane action is \cite{Bergshoeff:1987qx} (see also \cite{Hartnoll:2002th,Becker:2007zj}): 
\begin{eqnarray}\label{2}
S=\frac{1}{(2\pi)^2l_{11}^3}\int d^3\sigma&&\Big(\frac{(-\gamma)^{1/2}}{2}[\gamma^{ij}\partial_i X^{M}\partial_j X^{N} G_{MN}(X)-1] \cr
&&+\, \epsilon^{ijk}\partial_i X^{M}\partial_j X^{N}\partial_k X^{P}A_{MNP}(X)\Big)
\end{eqnarray}
where $i,j=0,1,2$ span the world-volume of the supermembrane with metric $\gamma^{ij}$, $M,N,P=0,...10$ are the space-time indices with metric $G_{MN}$ for space-time coordinates $X^{M}$, and $A_{MNP}(X)$ is a three-form field with strength $F=dA$. The constant $l_{11}$ sets the scale of the supermembrane. 

The equations of motion that follow from the action (\ref{2}) read:
\begin{eqnarray}\label{3}
\gamma_{ij} &=&\partial_i X^{M}\partial_j X^{N}G_{MN}(X)
\\
\partial_i\left((-\gamma)^{1/2}\gamma^{ij}\partial_j X^{P}\right)&=&
-(-\gamma)^{1/2}\gamma^{ij}\partial_i
X^{M}\partial_j X^{N}\Gamma^{P}_{MN}(X) \cr
&& -\epsilon^{ijk}\partial_i X^{M}\partial_j X^{N}\partial_k X^{Q}F^{P}_{MNQ}(X)\label{4}
\end{eqnarray} 

\noindent 
From eq. (\ref{3}) we obtain $\gamma^{ij}\partial_i X^{M}\partial_j X^{N}
G_{MN}(X)=\gamma^{ij}\gamma_{ij}=trI=3$, then the action, eq. (\ref{2}), becomes:
\begin{equation}\label{5}
 S=\frac{1}{(2\pi)^2l_{11}^3}\int d^3\sigma\left((-\gamma)^{1/2} +\epsilon^{ijk}\partial_i X^{M}\partial_j X^{N}\partial_k X^{P}A_{MNP}(X)\right)
\end{equation}

In order to apply the MRY method for calculating Wilson loops, we need to work with strings or string-like objects, so we have to pick a configuration of the membrane which has this shape. One way to do this is to consider a  compactification (\textit{\`a la Kaluza-Klein}) of third world-volume coordinate
on the 11-dimension of the background space. In this sense,
after taking the radio of compactification $R$ to zero, one
reduces the membrane to a string-like object. 
In the next sections we are going to discuss different backgrounds where we will implement this compactification.


\section{D=11 Elementary/electric 2-brane background}
The M2 brane  solution  corresponding to $N$ coincident M2-branes is given by the $SO(2,1)\times SO(8)$ symmetric geometry \cite{Duff:1990xz} 
(for a review see \cite{Review1,Review2}):
\begin{equation}\label{9}
ds^2=H^{-2/3}dx^{\mu}dx^{\nu}\eta_{\mu\nu}+H^{1/3}(dr^2+r^2d\Omega_7^2)
\end{equation} 
and the 3-form potential:
\begin{equation}\label{10}
A_{\mu\nu\lambda}=\epsilon_{\mu\nu\lambda}H^{-1},
\end{equation}
where the horizon function is defined as 
\begin{equation}\label{H}
H=1+(\dfrac{r_2}{r})^6\,,
\end{equation}
$r_2$ is a constant radius given by $(r_2)^6=32\pi^2N (l_{11})^6$ and $\mu,\nu,\lambda=0,1,2$ and others $A_{MNP}$ components are zero. 
The $S^7$ sphere metric can be written as:
$$d\Omega_7^2=\sin^2\theta_7\sin^2\theta_6\sin^2\theta_5\sin^2\theta_4\sin^2\theta_3\sin^2\theta_2d\theta_1^2+
\sin^2\theta_7\sin^2\theta_6\sin^2\theta_5\sin^2\theta_4\sin^2\theta_3d\theta_2^2$$
$$+\sin^2\theta_7\sin^2\theta_6\sin^2\theta_5\sin^2\theta_4d\theta_3^2+\sin^2\theta_7\sin^2\theta_6\sin^2\theta_5d\theta_4^2
+\sin^2\theta_7\sin^2\theta_6d\theta_5^2$$
\begin{equation}\label{11}
+\sin^2\theta_7d\theta_6^2+d\theta_7^2
\end{equation}

Another important feature of our calculation in common with the MRY method is to consider a static configuration to calculate the Wilson loop. We discuss this in the next subsection. 


\subsection{Static configuration}

We want the membrane coordinates to be in a static gauge so we set $X^0=\tau$, $X^1=x=\sigma$ and $X^2=0$,
also we want this membrane to be spread along the radial coordinate so we set $X^3=r=r(\sigma)$.

In order to do a dimensional world-volume reduction from the membrane to a string-like world-sheet, 
we must wrap the membrane along a $U(1)$ in the sphere. This can be done for instance with the choice 
$X^{4}=\theta_1=\delta/R$ 
and
$\theta_2=\theta_3=\theta_4=\theta_5=\theta_6=\theta_7=\dfrac{\pi}{2}$. 

Within this configuration and in the  M2 background geometry, one can  compute the action (\ref{5}). To this end, we first compute the 
$\gamma_{ij}$ matrix, which turns out  to be diagonal: 
$\gamma_{ij}=\text{diag}[-H^{-2/3},H^{-2/3}+(r')^2 H^{1/3},\frac{r^2}{R^2}H^{1/3}]$, where $r'=dr/dx$. Also, within this configuration, the membrane
decouples from the background 3-form field $A_{[3]}$. As a result, the action (\ref{5}) reduces to:
\begin{equation}\label{12}
 S=\frac{T}{(2\pi)l_{11}^ 3}\int dx\sqrt{\frac{r^2}{H}+r^ 2(r')^2}\,,
\end{equation}
where we used  $\int d\tau=T$ and $\int d\delta=R\int\theta_1=2\pi R$. Note also that the compactification radius $R$ has been cancelled out, so this action ($S$) is the same for the limiting case $R\rightarrow0$. 
From this action, the Lagrangian would be
\begin{equation}
 \mathcal{L}(r,r')=\frac{1}{(2\pi)l_{11}^ 3}\sqrt{\frac{r^2}{H}+r^ 2(r')^2}
 \end{equation}
Since it does not depend explicitly on $x$, there is a constant of motion
\begin{equation}\label{12a}
 \frac{r^2/H}{\sqrt{r^2/H+r^2(r')^2}}=\text{constant}
\end{equation}
Using symmetry arguments and defining $r_0$ to be the minimun value of $r$, which we supposed to occur at $x=0$, we find that the length $L$ of the string-like M2 membrane configuration can be written as  
\begin{equation}\label{13}
L=2 r_0\int_{1}^{r/r_0}dy\frac{\left(1+\frac{\epsilon}{y^6}\right)y^3}{\sqrt{y^8-y^6+\epsilon(y^8-1)}}
\end{equation}
and we defined $\epsilon=(r_2/r_0)^6$. 

Furthermore, the  energy of the static M2-brane configuration can  obtained from (\ref{12}) and (\ref{12a}) as:
\begin{equation}\label{14}
 {E}=\frac{r_0^2\sqrt{1+\epsilon}}{\pi l_{11}^3}\int_1^{\Lambda}dy\frac{y^5}{\sqrt{y^8-y^6+\epsilon(y^8-1)}}-2m_q,
\end{equation}
where $m_q$ is some constant that we will identify with the (infinite) mass of the quark in the dual theory when it exists and $\Lambda=r_1/r_0$, being $r_1$ the upper limit of coordinate $r$. In the following subsections, we are going to discuss some limiting cases for the shape of the string-like membrane and calculate explicitly the energy of the configuration with respect to the string-like length. These limiting cases correspond to different values of the parameters $\Lambda$ and $\epsilon$. 


\subsection{String far from the branes}
This case corresponds to the regime $r_2<<r_0$ ($\epsilon<<1$), 
i.e. the string-like membrane is placed very far from the M2-branes.
The string-like length and energy, given by expressions (\ref{13}) and (\ref{14}), then become:
\begin{eqnarray}
 L=2r_0\log(\sqrt{\Lambda^2-1}+\Lambda)
 +{r_0\epsilon}\Big[-\frac{(\Lambda^2-1)^{1/2}}{\Lambda}&-&\frac{(\Lambda^2-1)^{3/2}}{3\Lambda^3}+\frac{(\Lambda^2-1)^{5/2}}{5\Lambda^5}\cr\cr
 &&-\log(\sqrt{\Lambda^2-1}+\Lambda)\Big]\label{15}
\end{eqnarray}
\begin{eqnarray}
 E=\frac{r_0^2}{2\pi l_{11}^3}\Big[\log(\sqrt{\Lambda^2-1}+\Lambda)-{\Lambda^2}+{1}+{\epsilon}\Big(&-&\log(\sqrt{\Lambda^2-1}+\Lambda)\cr\cr
 &&+2\frac{(\Lambda^2-1)^{1/2}}{\Lambda}+\frac{(\Lambda^2-1)^{3/2}}{3\Lambda^3}\Big)\Big]-2m_q
 \cr\cr && \label{16}
\end{eqnarray}
These expressions for the length and energy of the string-like membrane can be further simplified choosing its relative size compared to the distance from the M2-branes $r_0$. We will discuss these cases in the following subsections.


\subsubsection{Long string}
 This subcase corresponds to $r_1>>r_0$ which means $\Lambda>>1$. In this regime, the length and energy of the string-like membrane, formulas (\ref{15}) and (\ref{16}) reduce to
\begin{equation}
 L=r_0\left(2-{\epsilon}\right)\log\Lambda
\end{equation}
\begin{equation}
 E=\frac{r_0^2}{2\pi l^3_{11}}\left[(1-\epsilon)\log\Lambda-\Lambda^2\right]-2m_q\,.
\end{equation}
From these last two relations we obtain
\begin{equation}
 E=\frac{r_0^2}{2\pi l_{11}^3}\left[\left(\frac{1}{2}-\frac{\epsilon}{4}\right)\frac{L}{ r_0}-e^{\frac{L}{r_0}(1+\epsilon/2)}\right]-2m_q
\end{equation}
This potential energy has a confining term proportional to the string-like length $L$ but is dominated by the exponential and then has a nonconfining behavior.


\subsubsection{Short string}
In this subcase the short string-like membrane corresponds to $r_1\sim r_0$, i.e  $\Lambda=1+\eta$, where $\eta$ is a small parameter ($\eta<<1$). In this regime, the length and energy of the string, formulas (\ref{15}) and (\ref{16}) read
\begin{equation}
L=2r_0(1-\epsilon)\sqrt{2\eta}
\end{equation}
\begin{equation}
E= \frac{r_0^2}{2\pi l_{11}^3}(1+\epsilon)\sqrt{2\eta}-2m_q
\end{equation}
As a result of these last two expressions, we have
\begin{equation}
 E=\frac{r_0}{4\pi l_{11}^3}(1+2\epsilon)L-2m_q
\end{equation}
This is a confining potencial since the energy grows with $L$.


\subsection{String close to the brane}
Now, we are going to examine the situation where the string-like membrane is located close to the M2-branes. This case corresponds to $r_2>>r_0$, which means $\epsilon>>1$, where $\epsilon=(r_2/r_0)^6$. This is nothing but  the $AdS_4\times S^7$ geometry.
Consequently, the length and energy of the string-like configuration, expressions (\ref{13}) and (\ref{14}),  become:
\begin{equation}\label{i}
L=2r_0\sqrt{\epsilon}\int_1^{\Lambda}dy\frac{1}{y^3\sqrt{y^8-1}}
\end{equation}
\begin{equation}\label{ii}
 E=\frac{r_0^2}{\pi l_{11}^3}\int_{1}^{\Lambda}dy\frac{y^5}{\sqrt{y^8-1}}-2m_q
\end{equation}
In the following, we are going to discuss the long and short string-like configurations for this background.


\subsubsection{Long string}
A long string-like membrane corresponds to $r_1>>r_0$ which means $\Lambda>>1$, where $\Lambda=r_1/r_0$. As a result, the length and energy of the string-like membrane given by Eqs. (\ref{i}) and (\ref{ii}) become:
\begin{equation}
L=2r_0\sqrt{\epsilon}\left(\int_{1}^{\infty}dy\frac{1}{y^3\sqrt{y^8-1}}+O(\Lambda^{-6})\right)=\frac{r_2^3}{r_0^2}\frac{\Gamma^2(\frac{1}{4})}{48\sqrt{2\pi}}\,,
\end{equation}
where we used the definition $\epsilon= (r_2/r_0)^6$, and 
\begin{equation}
{E}=\frac{r^2_0}{\pi l^3_{11}}\left[\int_{1}^{\infty}dy\left(\frac{y^5}{\sqrt{y^8-1}}-\frac{y}{4}\right)-\frac{1}{8}\right]=-
\frac{r_0^2}{\pi l^3_{11}}\frac{(2\pi)^{3/2}}{\Gamma^2(\frac{1}{4})}\,.
\end{equation}
This last formula needed to be regularized subtracting an infinite term that can be identified with the quark mass.
From these last expressions we finally obtain
\begin{equation}
 E=-\frac{r_2^3}{24 l_{11}^3}\frac{1}{L}\,.
\end{equation}
This non-confining potential energy corresponds to the conformal field theory result consistent with the string-like membrane placed in the $AdS_4\times S^7$ geometry. 


\subsubsection{Short string}
A short string-like membrane configuration corresponds to $r_1\sim r_0$, i.e  $\Lambda=1+\eta$, where $\eta<<1$. In this regime the length and energy of the configuration, formulas (\ref{i}) and (\ref{ii}) reduce to
\begin{equation}
 L= r_0\frac{\sqrt{\epsilon}}{\sqrt{2}}\sqrt{\eta}
\end{equation}
\begin{equation}
 E=\frac{r_0^2}{2\pi l_{11}^3\sqrt{2}}\sqrt{\eta}-2m_q
\end{equation}
Combining these expressions, and using the definition $\epsilon= (r_2/r_0)^6$, we find
\begin{equation}
 E=\frac{r_0^4}{2\pi l_{11}^3r_2^3}L-2m_q
\end{equation}
which corresponds to a linear confining potencial.


\section{D = 11 Solitonic/magnetic 5-brane background}
This is the case of $N$ M5-branes solitonic solution of supergravity, whose geometry is given by (for a review,
see for instance \cite{Review1,Review2}):
\begin{equation}
 ds^2=f^{-1/3}dx^{\mu}dx^{\nu}\eta_{\mu\nu}+f^{2/3}(dr^2+r^2d\Omega^2_4),
\end{equation}
where $\eta_{\mu\nu}$ is the mostly plus six-dimensional  Minkowski metric, $f$ is the horizon function given by
\begin{equation}\label{f}
 f=1+\left(\frac{r_5}{r}\right)^3,
\end{equation}
where $(r_5)^3= N \pi (l_{11})^3$, and the $S^4$ metric can be written as:
\begin{equation}
d\Omega_4^2=\sin^2\theta_4\sin^2\theta_3\sin^2\theta_2d\theta_1^2+\sin^2\theta_4\sin^2\theta_3d\theta_2^2
+\sin^2\theta_4d\theta_3^2+d\theta_4^2.
 \end{equation}
 

\subsection{Static configuration}
In analogy with the M2-brane space case, here for the M5-brane space we choose a static configuration
setting $X^0=\tau$, $X^1=\sigma\equiv x$,  $X^2=X^3=X^4=X^5=0$,
$X^6\equiv r=r(x)$. The string-like compactification is determined by the choice $X^7 \equiv\theta_1=\delta/R$ and $X^8=X^9=X^{10}=\theta_2=\theta_3=\theta_4=\pi/2$, where $R$ is the compactification radius. 
Consequently, in this configuration we have that the induced metric $ \gamma_{ij}=\partial_i X^{M}\partial_j X^{N}G_{MN}(X)$ reads 
\begin{equation}
 \gamma_{ij}=\text{diag}\left[-f^{-1/3},f^{-1/3}+(r')^2 f^{2/3},(\frac{r}{R})^2f^{2/3}\right]
\end{equation}
Also in this configuration $A_{[3]}$ decouples from the membrane, so action (\ref{5}) becomes
\begin{equation}
 S=\frac{T}{2\pi l_{11}^3}\int dx\sqrt{r^2+(r')^2r^2f}
\end{equation}
where, as in the M2 case, the integrations  $\int d\tau=T$ and 
$\int d\delta=R\int\theta_1=2\pi R$ have been done in this formula. 
Also, the compactification radius $R$ has been cancelled out, so this action $S$ is equal to the string-like limit case $R\rightarrow0$.

Since the Lagrangian corresponding to the action $S$  does not depend explicitly on $x$, the solution
satisfies:
\begin{equation}
 \frac{r^2}{\sqrt{r^2+(r')^2r^2 f}}=\text{constant}
\end{equation}
It is clear that this equation has a minimum $r_0$ which, by symmetry arguments, occurs at $x=0$, so the length of the string-like membrane is:
\begin{equation}\label{17}
 L=2 r_0\int_{1}^{r/r_0}dy\frac{(1+\epsilon/y^3)^{1/2}}{(y^2-1)^{1/2}}
\end{equation}
where we defined the parameter $\epsilon=({r_5}/{r_0})^3$.

The corresponding energy is given by:
\begin{equation}\label{18}
 E=\frac{r_0^2}{\pi l_{11}^3}\int_1^{r/r_0}dy\frac{y^2(1+\epsilon/y^3)^{1/2}}{(y^2-1)^{1/2}}-2m_q
\end{equation}
where $m_q$ is a possibly infinite term which can be identified with a very large quark mass in the dual gauge theory when this theory exists.


\subsection{String close to the brane}
In this regime $r_5>>r_0$ ($\epsilon>>1$) so we are analyzing the case where the string is placed very close to the brane.
Also this case corresponds to the $AdS_7\times S^4$ geometry, then eqs (\ref{17}) and (\ref{18}) become
\begin{equation}\label{19}
 L=2 r_0\epsilon^{1/2}\int_{1}^{\Lambda}dy\frac{1}{y^{3/2}(y^2-1)^{1/2}}
\end{equation}
\begin{equation}\label{20}
  E=\frac{r_0^2}{\pi l_{11}^3}\sqrt{\epsilon}\int_1^{\Lambda}dy\frac{y^{1/2}}{(y^2-1)^{1/2}}-2m_q
\end{equation}
where $\Lambda=r_1/r_0$ and $r_1$ is the position of the ends of the string-like membrane. 
Next, we consider different limits on $\epsilon$ and $\Lambda$ and calculate the corresponding string-like length and energy.


\subsubsection{Long string}
In this case $r_1>>r_0$ ($\Lambda>>1$), then the string-like membrane length and energy, given by eqs.  (\ref{19}) and  (\ref{20}), become:
\begin{equation}
L=2 r_0\epsilon^{1/2}\int_{1}^{\infty}dy\frac{1}{y^{3/2}(y^2-1)^{1/2}}=
2 \frac{ r_5^{3/2}}{r_0^{1/2}}\, c \,,
\end{equation}
 where we used that $\epsilon= (r_5/r_0)^3$, $c={(2\pi)^{3/2}}/{\Gamma^2(\frac{1}{4})}$, and 
\begin{equation}
 E=\frac{r_0^2}{\pi l_{11}^3}\left[\int_1^{\infty}dy
 \left(\frac{y^{1/2}}{(y^2-1)^{1/2}}-\frac{y^{-1/2}}{4}\right)-\frac{1}{2}\right]=
 -\frac{r_5^{3/2}r_0^{1/2}c}{\pi l_{11}^3}\,.
\end{equation}
Consequently, the energy of the string-like configuration can be written as
\begin{equation}
 E=-\frac{r^3_5c^2}{\pi l_{11}^3}\frac 2 L\,. 
\end{equation}
This non-confining energy corresponds to the conformal field theory result of the dual theory. 
This is the expected result since for the string close to the brane the geometry is the $AdS_7\times S^4$. 


\subsubsection{Short string}
In this regime $r_1\sim r_0 $, i.e. $\Lambda=1+\eta$ where $\eta<<1$. 
Then, the length and energy of the string-like configuration, eqs. (\ref{19}) and (\ref{20}), become:
\begin{equation}
 L=\frac{r_5^{3/2}}{r_0^{1/2}}\sqrt{2\eta}\, [1+O(\eta)]
\end{equation}
\begin{equation}
 E=\frac{r_0^{1/2}r_5^{3/2}}{2\pi l_{11}^3}\sqrt{2\eta}\, [1+O(\eta)]-2m_q
\end{equation}
Consequently the energy of the configuration is given by 
\begin{equation}
 E=\frac{r_0}{2\pi l_{11}^3}L-2m_q\,.
\end{equation}
This corresponds to a linear confining potential in the dual theory. 


\subsection{String far from the brane}
In this case $r_5<<r_0$ ($\epsilon<<1$) so the length and energy of the string-like configuration, eqs. (\ref{17}) and (\ref{18}) are given by
\begin{equation}\label{21}
 L=2 r_0\log(\Lambda+\sqrt{\Lambda^2-1})+
 {r_0\epsilon}{}\left(\frac{\sqrt{\Lambda^2-1}}{2\Lambda^2}+\arccos\frac{1}{\Lambda}\right)
\end{equation}
\begin{equation}\label{22}
 E=\frac{r_0^2}{2\pi l_{11}^3}\left[\log(\Lambda+\sqrt{\Lambda^2-1})-\Lambda^2+1
 +\epsilon\arccos\frac{1}{\Lambda}\right]-2m_q
\end{equation}
where $\epsilon=(r_5/r_0)^3$.  In the following, we discuss particular limits on $\Lambda= r_1/r_0$. 


\subsubsection{Long string}
In this case $r_1>>r_0$ ($\Lambda>>1$) so eqs. (\ref{21}) and (\ref{22}) become:
\begin{equation}
 L=2 r_0\log\Lambda
\end{equation}
\begin{equation}
 E=\frac{r_0^2}{2\pi l_{11}^3}[\log\Lambda-\Lambda^2]-2m_q
\end{equation}
Consequently, the energy of the configuration is given by
\begin{equation}
 E=\frac{r_0^2}{2\pi l_{11}^3}\left[\frac{L}{2 r_0}-e^{L/r_0}\right]-2m_q
\end{equation}
This energy has a linear term but is dominated by the exponential term so that it corresponds to a non-confining potential.


\subsubsection{Short string}
In this regime $r_1\sim r_0$ i.e. $\Lambda=1+\eta$ where $\eta<<1$, so eqs. (\ref{21}) and (\ref{22}) become:
\begin{equation}
 L={r_0}\left(1+\frac{\epsilon}{2}\right)\sqrt{2\eta}\, [1+O(\eta)]
\end{equation}
\begin{equation}
 E=\frac{r_0^2}{2\pi l_{11}^3}\left(1+\frac{\epsilon}{2}\right)\sqrt{2\eta}\, [1+O(\eta)]-2m_q
\end{equation}
Consequently, the configuration energy in this case is 
\begin{equation}
 E=\frac{r_0}{2\pi l_{11}^3}L-2m_q
\end{equation}
This corresponds to a linear confining potential.


\section{Conclusions}

With a membrane setup we have studied Wilsons loops in M2 and M5 spaces, mainly obtaining classical potential interactions which may be interpreted as the interaction of a quark-antiquark pair at least when the M2 and M5 brane spaces reduce to the $AdS_4\times S^7$ and $AdS_7\times S^4$ spaces, respectively. In these cases the dual theories are supersymmetric $SU(N)$ gauge theories, according to the AdS/CFT correspondence \cite{Itzhaki:1998dd}. 
We analyzed some regimes and in some cases a linear confinement was found (typically for short string-like membranes) while in other cases a non-confining Coulomb-like behavior was obtained (for long string-like membranes) in agreement with a conformal field theory result. It is interesting to note that this situation resembles the QCD behavior where confinement holds at low energies and a non-confining situation (asymptotic freedom) appear at high energies. A similar result \cite{henrique} was found in the study of Wilson loops in D3-brane space associated with the $AdS_5\times S^5$ limit. These results suggest a confinement/deconfinement phase transition. It would be interesting to perform a numerical study of these problems in order to understand the nature of these possible phase transitions. 

A comment is order here about the dual theories of our M2- and M5-brane backgrounds. In the near horizon approximation these backgrounds reduce to $AdS_4\times S^7$ and $AdS_7\times S^4$ spaces, as is well known. In these limitings cases the dual theories are supersymmetric $SU(N)$ gauge theories. The more subtle situations are the cases where we do not use the near horizon approximation. In these cases, the horizon functions, Eqs.  (\ref{H}) and (\ref{f}), imply a flat space limit far from the horizon. This is different from the AdS case where the curvature is constant. The flat space limit modifies the dual theory adding a higher dimensional irrelevant operator on the dual theory turning the problem more involved. Any way, we were able to calculate the configuration energy of the string-like membrane on the gravity side. Clearly, the irrelevant operator in the dual theory turns the interpretation of the duality more involved. See ref. \cite{Nunez:2009da} for a discussion. See also ref. \cite{Bea:2015fja}. 

In Refs. \cite{Gomis:2006sb,Gomis:2006im} Wilson loops in AdS5 were related to D3-branes. 
Based on these results we expect that analogous relations could be established for Wilson loops in AdS4 and AdS7 in relation to M2- and M5-branes.

Recently, a proposal to calculate 1/N corrections to the D3-brane Wilson loop was
given in \cite{Faraggi:2011bb,Buchbinder:2014nia,Faraggi:2014tna}. 
We expect that such an approach could also be applied to the case of M2-
and M5-brane Wilson loops.


\acknowledgments We would like to thank Carlos N\'u\~nez for many enlightening  discussions. We also thank interesting and useful correspondence with Mike Duff, Mir Faizal and Leo Pando Zayas. We thank also financial support from CNPq - Brazilian agency.




\begin{thebibliography}{99}

\bibitem{maldacena1} 
  J.~M.~Maldacena,
  ``The Large N limit of superconformal field theories and supergravity,''
  Int.\ J.\ Theor.\ Phys.\  {\bf 38}, 1113 (1999)
  [Adv.\ Theor.\ Math.\ Phys.\  {\bf 2}, 231 (1998)]
  [hep-th/9711200].

\bibitem{GKP} 
  S.~S.~Gubser, I.~R.~Klebanov and A.~M.~Polyakov,
  ``Gauge theory correlators from noncritical string theory,''
  Phys.\ Lett.\ B {\bf 428}, 105 (1998)
  [hep-th/9802109].
  
\bibitem{Witten1} 
  E.~Witten,
  ``Anti-de Sitter space and holography,''
  Adv.\ Theor.\ Math.\ Phys.\  {\bf 2}, 253 (1998)
  [hep-th/9802150].
  
\bibitem{Review1} 
  J.~L.~Petersen,
  ``Introduction to the Maldacena conjecture on AdS / CFT,''
  Int.\ J.\ Mod.\ Phys.\ A {\bf 14}, 3597 (1999)
  [hep-th/9902131].
  
\bibitem{Review2} 
  O.~Aharony, S.~S.~Gubser, J.~M.~Maldacena, H.~Ooguri and Y.~Oz,
  ``Large N field theories, string theory and gravity,''
  Phys.\ Rept.\  {\bf 323}, 183 (2000)
  [hep-th/9905111].

\bibitem{maldacena2}
J.M.Maldacena, 
``Wilson loops in Large-N field theories," 
{Phys. Rev. Lett.} {\bf 80}(1998) 4859 
[hep-th/9803002].

\bibitem{Rey:1998ik} 
  S.~J.~Rey and J.~T.~Yee,
  ``Macroscopic strings as heavy quarks in large N gauge theory and anti-de Sitter supergravity,''
  Eur.\ Phys.\ J.\ C {\bf 22}, 379 (2001)
  [hep-th/9803001].

\bibitem{kinar} 
  Y.~Kinar, E.~Schreiber and J.~Sonnenschein,
  ``Q anti-Q potential from strings in curved space-time: Classical results,''
  Nucl.\ Phys.\ B {\bf 566}, 103 (2000)
  [hep-th/9811192].

\bibitem{finiteT} 
  A.~Brandhuber, N.~Itzhaki, J.~Sonnenschein and S.~Yankielowicz,
  ``Wilson loops in the large N limit at finite temperature,''
  Phys.\ Lett.\ B {\bf 434}, 36 (1998)
  [hep-th/9803137].
  
\bibitem{brandhuber} A.Brandhuber, N. Itzhaki, J.Sonnenschein and S. Yankielowicz 
{``Wilson Loops, Confinement and Phase Transitions in Large $N$ Gauge Theories from Supergravity}," 
{JHEP} {\bf 06} (1998) 001 [hep-th/9803263].

\bibitem{Witten2} 
  E.~Witten,
  ``Anti-de Sitter space, thermal phase transition, and confinement in gauge theories,''
  Adv.\ Theor.\ Math.\ Phys.\  {\bf 2}, 505 (1998)
  [hep-th/9803131].
  
\bibitem{henrique}  
H.~Boschi-Filho and N.~R.~F.~Braga,
  ``Wilson loops for a quark anti-quark pair in D3-brane space,''
  JHEP {\bf 0503}, 051 (2005)
  [hep-th/0411135].

\bibitem{henrique2} 
  H.~Boschi-Filho, N.~R.~F.~Braga and C.~N.~Ferreira,
  ``Static strings in Randall-Sundrum scenarios and the quark anti-quark potential,''
  Phys.\ Rev.\ D {\bf 73}, 106006 (2006)
  [Erratum-ibid.\ D {\bf 74}, 089903 (2006)]
  [hep-th/0512295].

\bibitem{Andreev:2006ct} 
  O.~Andreev and V.~I.~Zakharov,
  ``Heavy-quark potentials and AdS/QCD,''
  Phys.\ Rev.\ D {\bf 74}, 025023 (2006)
  [hep-ph/0604204].

\bibitem{henrique3} 
  H.~Boschi-Filho, N.~R.~F.~Braga and C.~N.~Ferreira,
  ``Heavy quark potential at finite temperature from gauge/string duality,''
  Phys.\ Rev.\ D {\bf 74}, 086001 (2006)
  [hep-th/0607038].
  
\bibitem{HP} 
  S.~W.~Hawking and D.~N.~Page,
  ``Thermodynamics of Black Holes in anti-De Sitter Space,''
  Commun.\ Math.\ Phys.\  {\bf 87}, 577 (1983).
  
\bibitem{Andreev:2006eh} 
  O.~Andreev and V.~I.~Zakharov,
  ``The Spatial String Tension, Thermal Phase Transition, and AdS/QCD,''
  Phys.\ Lett.\ B {\bf 645}, 437 (2007)
  [hep-ph/0607026].
  
\bibitem{Herzog} 
  C.~P.~Herzog,
  ``A Holographic Prediction of the Deconfinement Temperature,''
  Phys.\ Rev.\ Lett.\  {\bf 98}, 091601 (2007)
  [hep-th/0608151].
  
  \bibitem{Kajantie:2006hv} 
  K.~Kajantie, T.~Tahkokallio and J.~T.~Yee,
  ``Thermodynamics of AdS/QCD,''
  JHEP {\bf 0701}, 019 (2007)
  [hep-ph/0609254].
  
\bibitem{Andreev:2006nw} 
  O.~Andreev and V.~I.~Zakharov,
  ``On Heavy-Quark Free Energies, Entropies, Polyakov Loop, and AdS/QCD,''
  JHEP {\bf 0704}, 100 (2007)
  [hep-ph/0611304].

\bibitem{henrique4} 
  C.~A.~Ballon Bayona, H.~Boschi-Filho, N.~R.~F.~Braga and L.~A.~Pando Zayas,
  ``On a Holographic Model for Confinement/Deconfinement,''
  Phys.\ Rev.\ D {\bf 77}, 046002 (2008)
  [arXiv:0705.1529 [hep-th]].

 \bibitem{duff}
M.J.Duff, P.S.Howe,T.Inami and K.S.stelle, 
``Superstrings in D=10 from supermenbranes in D=11" 
{Phys. Lett.B} {\bf 191}(1987).

\bibitem{Farquet:2014bda} 
  D.~Farquet and J.~Sparks,
  ``Wilson loops on three-manifolds and their M2-brane duals,''
  JHEP {\bf 1412}, 173 (2014)
  [arXiv:1406.2493 [hep-th]].
  
\bibitem{Hartnoll:2009sz} 
  S.~A.~Hartnoll,
  ``Lectures on holographic methods for condensed matter physics,''
  Class.\ Quant.\ Grav.\  {\bf 26}, 224002 (2009)
  [arXiv:0903.3246 [hep-th]].
  
\bibitem{Herzog:2009xv} 
  C.~P.~Herzog,
  ``Lectures on Holographic Superfluidity and Superconductivity,''
  J.\ Phys.\ A {\bf 42}, 343001 (2009)
  [arXiv:0904.1975 [hep-th]].
  
\bibitem{McGreevy:2009xe} 
  J.~McGreevy,
  ``Holographic duality with a view toward many-body physics,''
  Adv.\ High Energy Phys.\  {\bf 2010}, 723105 (2010)
  [arXiv:0909.0518 [hep-th]].
  
\bibitem{Horowitz:2010gk} 
  G.~T.~Horowitz,
  ``Introduction to Holographic Superconductors,''
  Lect.\ Notes Phys.\  {\bf 828}, 313 (2011)
  [arXiv:1002.1722 [hep-th]].
  
\bibitem{Seiberg:1996vs} 
  N.~Seiberg and E.~Witten,
  ``Comments on string dynamics in six-dimensions,''
  Nucl.\ Phys.\ B {\bf 471}, 121 (1996)
  [hep-th/9603003].
  
\bibitem{Witten:1997kz} 
  E.~Witten,
  ``New 'gauge' theories in six-dimensions,''
  JHEP {\bf 9801}, 001 (1998)
  [Adv.\ Theor.\ Math.\ Phys.\  {\bf 2}, 61 (1998)]
  [hep-th/9710065].
  
\bibitem{Bergshoeff:1987qx} 
  E.~Bergshoeff, E.~Sezgin and P.~K.~Townsend,
  ``Properties of the Eleven-Dimensional Super Membrane Theory,''
  Annals Phys.\  {\bf 185}, 330 (1988).

\bibitem{Hartnoll:2002th} 
  S.~A.~Hartnoll and C.~Nunez,
  ``Rotating membranes on G(2) manifolds, logarithmic anomalous dimensions and N=1 duality,''
  JHEP {\bf 0302}, 049 (2003)
  [hep-th/0210218].
  
\bibitem{Becker:2007zj} 
  K.~Becker, M.~Becker and J.~H.~Schwarz,
  ``String theory and M-theory: A modern introduction,''
  Cambridge, UK: Cambridge Univ. Pr. (2007) 739 p

\bibitem{Duff:1990xz} 
  M.~J.~Duff and K.~S.~Stelle,
  ``Multimembrane solutions of D = 11 supergravity,''
  Phys.\ Lett.\ B {\bf 253}, 113 (1991).

\bibitem{Itzhaki:1998dd} 
  N.~Itzhaki, J.~M.~Maldacena, J.~Sonnenschein and S.~Yankielowicz,
  ``Supergravity and the large N limit of theories with sixteen supercharges,''
  Phys.\ Rev.\ D {\bf 58}, 046004 (1998)
  [hep-th/9802042].

\bibitem{Nunez:2009da} 
  C.~Nunez, M.~Piai and A.~Rago,
  ``Wilson Loops in string duals of Walking and Flavored Systems,''
  Phys.\ Rev.\ D {\bf 81}, 086001 (2010)
  [arXiv:0909.0748 [hep-th]].

\bibitem{Bea:2015fja} 
  Y.~Bea, J.~D.~Edelstein, G.~Itsios, K.~S.~Kooner, C.~Nunez, D.~Schofield and J.~A.~Sierra-Garcia,
  ``Compactifications of the Klebanov-Witten CFT and new AdS$_{3}$ backgrounds,''
  JHEP {\bf 1505}, 062 (2015)
  [arXiv:1503.07527 [hep-th]].

\bibitem{Gomis:2006sb} 
  J.~Gomis and F.~Passerini,
  ``Holographic Wilson Loops,''
  JHEP {\bf 0608}, 074 (2006)
  [hep-th/0604007].
  
\bibitem{Gomis:2006im} 
  J.~Gomis and F.~Passerini,
  ``Wilson Loops as D3-Branes,''
  JHEP {\bf 0701}, 097 (2007)
  [hep-th/0612022].
  
\bibitem{Faraggi:2011bb} 
  A.~Faraggi and L.~A.~Pando Zayas,
  ``The Spectrum of Excitations of Holographic Wilson Loops,''
  JHEP {\bf 1105}, 018 (2011)
  [arXiv:1101.5145 [hep-th]].
  
\bibitem{Buchbinder:2014nia} 
  E.~I.~Buchbinder and A.~A.~Tseytlin,
  ``1/N correction in the D3-brane description of a circular Wilson loop at strong coupling,''
  Phys.\ Rev.\ D {\bf 89}, no. 12, 126008 (2014)
  [arXiv:1404.4952 [hep-th]].

\bibitem{Faraggi:2014tna} 
  A.~Faraggi, J.~T.~Liu, L.~A.~Pando Zayas and G.~Zhang,
  ``One-loop structure of higher rank Wilson loops in AdS/CFT,''
  Phys.\ Lett.\ B {\bf 740}, 218 (2015)
  [arXiv:1409.3187 [hep-th]].



\end{thebibliography}
\end{document}